\documentclass[sigconf]{acmart}

\AtBeginDocument{%
  \fancyhead{}
}

\setcopyright{none}
\renewcommand\footnotetextcopyrightpermission[1]{}

\usepackage{balance}

\usepackage{threeparttable}
\usepackage{tikz}
\usepackage{graphicx}
\usepackage[linesnumbered,ruled,vlined]{algorithm2e}
\usepackage{booktabs}
\usepackage{listings}
\usepackage{algpseudocode}
\usepackage{smartdiagram}
\algrenewcommand\Call[2]{\textsc{#1}(#2)}   
\usepackage{amsmath}
\usepackage{amsthm}
\usepackage{tikzpagenodes}
\usepackage{tabularx}
\usepackage{pifont}
\usepackage{xspace}
\usepackage[normalem]{ulem}
\usepackage{subfigure}
\usepackage{wasysym}
\usepackage{subcaption}
\usepackage{url}
\usepackage[dvipsnames]{xcolor}
\usepackage{hyperref}   
\usepackage{tcolorbox}
\usepackage{comment}
\tcbuselibrary{breakable}

\newcommand{\pmark}{\LEFTcircle}
\newcommand{\cmark}{\ding{51}}%
\newcommand{\xmark}{\ding{55}}%

\newcommand{\BfPara}[1]{{\noindent\textbf{#1.}}\xspace}

\title{Toward Intelligent Networks via AI-Aware GPU-Native Packet Processing}

\author{Seyed Mohammad Mehdi Mirnajafizadeh}
\email{m.mirnajafi@wayne.edu}
\affiliation{%
  \institution{Wayne State University}
  \city{Detroit}
  \state{Michigan}
  \country{USA}
}

\author{Yiwen Hu}
\email{huyiwen@umbc.edu}
\affiliation{%
  \institution{University of Maryland, Baltimore County}
 \city{Baltimore}
  \state{Maryland}
  \country{USA}}

\author{Rhongho Jang}
 \email{r.jang@wayne.edu}
\affiliation{%
  \institution{Wayne State University}
  \city{Detroit}
  \state{Michigan}
  \country{USA}}
  
\begin{abstract}
The integration of inline Artificial Intelligence (AI) models into critical network infrastructure is fundamentally bottlenecked by the high latency and synchronization overhead of CPU-mediated packet processing. While legacy GPU offload and recent CPU-bypass frameworks attempt to bridge this gap, they remain trapped in proprietary ecosystems or still rely on the host CPU and coarse-grained batching to coordinate stateful telemetry and AI pipeline execution. In this paper, we propose AGP, a novel framework that promotes the GPU from a passive accelerator to a primary data-path controller, answering what changes architecturally when the GPU autonomously owns the complete packet-to-inference pipeline. By enabling GPU-native packet processing, contention-free in-GPU stateful aggregation, and a persistent mega-kernel for continuous AI inference, AGP removes the CPU from the critical path. Our evaluation demonstrates that native in-GPU packet processing sustains line-rate throughput while being over $6.1\times$ more power-efficient. Furthermore, our integrated Intrusion Detection System (IDS) stress test eliminates the legacy CPU-GPU synchronization tax, reducing end-to-end latency by $7.9\times$ (up to $35\times$ p99) and accelerating whole-system inference throughput by $9.3\times$ using only $\sim$2\% of the GPU's thread capacity.
\end{abstract}

\begin{document}




\maketitle
\section{Introduction}

Despite rapid integration of inline Artificial Intelligence (AI) into critical network infrastructure for applications such as telemetry~\cite{Jinetal2023, Zhengetal2024, SrivastavaSekar2024, Kapooretal2025}, infrastructure provisioning~\cite{Chenetal2018, Zhangetal2020}, and intrusion detection systems~\cite{mirnajafizadeh2026defeating, kim2025sketchfeature, mirnajafizadeh2024enhancing}, the deployment of these systems remains constrained by a legacy architectural paradigm: treating the GPU strictly as a passive, host-dependent coprocessor. 

Over the past two decades, both classic GPU packet-processing frameworks~\cite{Hanetal2010, Vasiliadisetal2011, Jamshedetal2012, Kimetal2014, Kimetal2015} and modern AI-integrated network systems have treated packet processing and GPU execution as two isolated pipelines~\cite{mirnajafizadeh2026defeating, fu2024detecting, fu2023detecting, kim2025sketchfeature}. Whether packets are polled by kernel-bypass drivers~\cite{dpdk} or filtered by in-kernel fast paths~\cite{hoiland2018xdp, vieira2020ebpf}, data and control still traverse a slow, host-mediated bridge: features are collected in the data plane, coordinated by CPUs, and pushed across PCIe in batches to amortize launch overheads. This twisted path introduces prohibitive latency, power overhead, and poor performance.

To address the inefficiency of this CPU-mediated bridge, the networking community has recently embraced CPU-bypass frameworks. Technologies such as GPUDirect RDMA \cite{NVIDIA2020GPUDirectRDMA} and NVIDIA GPUNetIO~\cite{nvidia_gpunetio} allow the Network Interface Card (NIC) to write directly to GPU memory and offload basic queue interactions. However, we argue that these I/O-layer optimizations create an
\textit{illusion} of GPU autonomy: they stop at packet delivery and leave state
management and inference scheduling on the host.

Closing this gap within the GPU exposes three technical challenges that prior work leaves open. \textbf{(C1) Owning NIC queues on an open stack.} DPDK poll-mode
drivers~\cite{dpdk} assume queues reside in host memory, and legacy GPU
frameworks~\cite{Hanetal2010, Kimetal2015} inherit this model by polling from
the CPU. While DPDK's gpudev library~\cite{dpdk_gpudev} places packet payloads in GPU memory, the CPU still polls the queues and relays packet addresses to
the GPU through a communication list. GPUNetIO~\cite{nvidia_gpunetio} moves packet reception onto the GPU, but only through a closed stack tied to recent NICs~\cite{CONNECTX7,
CONNECTX8}, so the queue layout cannot be co-designed with the state and
inference stages that follow. \textbf{(C2) Updating shared telemetry state under SIMT.} Prior GPU packet processors are stateless per batch~\cite{Hanetal2010, Kimetal2015}, while stateful telemetry is typically maintained on CPUs or switch ASICs~\cite{kim2025sketchfeature, mirnajafizadeh2024enhancing}, so features must be copied and reshaped into tensors before inference. However, moving this state onto the GPU introduces a new challenge: traffic bursts make many threads within a warp update the same flow, forcing their atomic operations to serialize, and heterogeneous headers cause threads to diverge during parsing. \textbf{(C3) Triggering inference without the host.} While CUDA Graphs~\cite{nvidia_cudagraphs} and recent mega-kernels~\cite{cheng2025mirage} reduce launch overhead, they still rely on the host to launch batches or dispatch requests. Together, these limitations raise a fundamental architectural question: \textit{what changes when the GPU is promoted from a passive I/O target to the autonomous owner of the entire network data path, and what new capabilities does this unlock?}

To answer this question, we argue for a fundamental shift in how accelerators are deployed: to unlock low-latency, AI-driven networking, the GPU must transition from a host-dependent coprocessor into an autonomous execution domain. We therefore propose an \textbf{AI-aware GPU-native Packet processing framework (AGP)}, which elevates the GPU into a fully autonomous network endpoint for inline AI inference.
Specifically, instead of relying on a host CPU or proprietary NIC firmware, AGP enables CUDA cores to autonomously manage low-level Work/Completion Queue Entries (WQEs/CQEs) in VRAM by extending the open DPDK stack~\cite{dpdk}. To bridge the gap between raw byte streams and AI tensors, AGP extracts and aggregates features in place as packets arrive, producing GPU-resident tensors that inference consumes without an intermediate copy. Moreover, packet I/O, feature extraction, and multi-layer model inference are fused into a persistent mega-kernel with device-side manager-worker scheduling, eliminating host kernel launches and batch-wait delays.

We developed an AGP prototype that implements a state-of-the-art Intrusion Detection System (IDS)~\cite{kim2025sketchfeature} as a demanding end-to-end stress test requiring line-rate ingestion, high-concurrency state updates, and continuous AI inference. Our benchmarks reveal that autonomous in-GPU packet processing delivers line-rate performance comparable to 16 dedicated polling cores, while being over 6.1$\times$ more power-efficient. Furthermore, we achieve 9.3$\times$ higher inference throughput than CPU-bypass frameworks using only $\sim$2\% of the GPU's thread capacity. These results answer our opening question. When the GPU owns the data path, (i) the unit of scheduling shrinks from a host-launched batch to a single tensor event, removing batch-wait latency; (ii) state lives where inference runs,
removing the byte-to-tensor copy; and (iii) the CPU leaves the fast path
entirely, freeing all host cores. This GPU efficiency as a network endpoint therefore unlocks new opportunities: 1) multi-tenant GPU service offloading and 2) a unified compiler for co-located network functions. We summarize the key contributions of this work as follows:

\begin{itemize}
\vspace{-1mm}

% \item \textbf{Decoupling GPU Data-Path Control from CPU:} We break away from proprietary constraints (e.g., DOCA/GPUNetIO) by extending open DPDK to enable queue management directly into GPU user space, eliminating host polling threads and CPU control syncs.

\item \textbf{Open GPU-Resident NIC Queues:} We extend open DPDK to place NIC
rings in VRAM and poll them from a persistent CUDA kernel, following the
GPUNetIO model but without its proprietary stack or NIC-generation limits.

\item \textbf{Eliminating the Byte-to-Tensor Transformation Overhead:} AGP embeds hardware-aligned stateful structures (e.g., probabilistic sketches) into the ingestion path, updated in place via warp-cooperative register reductions to avoid global atomic serialization.

\item \textbf{Device-Side Mega-Kernel AI Scheduling:} We replace host-driven, batched multi-kernel launches with a persistent mega-kernel and on-device manager-worker queues, triggering inference continuously in on-chip memory as individual tensors mature, eliminating CPU-GPU synchronization.

\item \textbf{End-to-End Validation and Co-Location Headroom:} We validate the AGP prototype on a 100 Gbps testbed and a stateful IDS stress test, demonstrating GPU data-path ownership frees all host CPU cores and over 70\% of GPU SMs for co-located applications.

\end{itemize}

\section{Background: AI Deployment Gap}

The pursuit of high-speed, AI-driven networking forces developers to navigate heavily fragmented hardware interfaces. While both classic GPU packet-processing frameworks and recent CPU-bypass technologies (e.g., GPUDirect RDMA and NVIDIA GPUNetIO) accelerate isolated stages of the pipeline, a deep architectural mismatch remains between line-rate network control and continuous AI execution.

\BfPara{Limitations of Host-Orchestrated Architecture} Historically, both software data planes and early GPU packet-processing frameworks rely on CPU orchestration.  In software data planes (e.g., DPDK~\cite{fu2023detecting, fu2024detecting, mirnajafizadeh2026defeating, dpdk, liu2025provguard}), the CPU constantly polls the NIC Work/Completion Queue Entries (WQEs/CQEs), parses packet headers, and batches data for the GPU. Pioneering GPU packet-processing systems~\cite{Hanetal2010, Vasiliadisetal2011, Jamshedetal2012, Kimetal2014, Kimetal2015} offloaded compute-heavy tasks to the GPU, but still relied on the host CPU to poll the NIC and accumulate massive packet batches in host DRAM before copying them across PCIe. Even with programmable switches~\cite{Tofino, Tofino2} where ASICs aggregate telemetry at line rate, the CPU must still retrieve, normalize, and push features across PCIe~\cite{mirnajafizadeh2024enhancing, kim2025sketchfeature, pan2026netradar}. Across all these paradigms, the CPU bears the burden of preparing the data and scheduling the AI inference via standard frameworks (e.g., TensorFlow, PyTorch). This host-driven scheduling introduces a severe synchronization cost, as standard deep learning frameworks operate by launching a sequence of dozens of discrete micro-kernels (e.g., convolution, ReLU, pooling) with a context-switching overhead of 5 to 10 microseconds per-kernel~\cite{kwon2020nimble, chen2018tvm}. This repeated movement of control and data across the PCIe bus introduces high latency that can quickly dominate the end-to-end inference pipeline. Additionally, the recent persistent mega-kernels (e.g., Mirage~\cite{cheng2025mirage}) that fuse model layers on the GPU still rely on the host CPU to feed requests and lack native network I/O support.

\begin{table}[t]
\centering
\caption{Current Network-to-AI Integration Architectures. (\pmark: with DPDK gpudev, payloads land in GPU memory, but the CPU still polls the queues
and relays packet addresses).}\vspace{-3mm}
\label{tab:architecture_comparison}
\resizebox{\columnwidth}{!}{%
\begin{tabular}{@{}l ccc cc@{}}
\toprule
 & \multicolumn{3}{c}{\textbf{Packet Processing}} & \multicolumn{2}{c}{\textbf{AI Integration}} \\
\cmidrule(lr){2-4} \cmidrule(lr){5-6}
\textbf{Approach} & 
\textbf{\shortstack{Development\\Flexibility}} & 
\textbf{\shortstack{Direct GPU\\I/O}} & 
\textbf{\shortstack{GPU Queue\\Control}} & 
\textbf{\shortstack{In-GPU\\Telemetry}} & 
\textbf{\shortstack{Device-Side\\AI Sched.}} \\
\midrule
Host-Mediated~\cite{dpdk, Hanetal2010} & \cmark & \pmark & \xmark & \xmark & \xmark \\
CPU-Bypass~\cite{nvidia_doca, nvidia_gpunetio} & \xmark & \cmark & \cmark & \xmark & \xmark \\
Mega-Kernel~\cite{cheng2025mirage} & \cmark & -- & -- & \xmark & \cmark \\
\midrule
\textbf{AGP (Ours)} & \textbf{\cmark} & \textbf{\cmark} & \textbf{\cmark} & \textbf{\cmark} & \textbf{\cmark} \\
\bottomrule
\end{tabular}%
}
\vspace{-3mm}
\end{table}

\BfPara{Limitations of Proprietary CPU-Bypass Frameworks} To mitigate host-coordination overhead, the industry has shifted toward CPU-bypass frameworks (e.g., DOCA~\cite{nvidia_doca}, GPUNetIO~\cite{nvidia_gpunetio}) and SmartNIC~\cite{nvidia_bluefield_smartnic} architectures. While these technologies successfully allow the NIC to write packets directly to GPU memory and offload tasks like doorbell ringing, they stop at the raw I/O boundary and leave three fundamental deployment gaps. First, the data delivery path operates as a closed, proprietary black box, restricted to the newest hardware generations (e.g., ConnectX-7~\cite{CONNECTX7} and newer~\cite{CONNECTX8}) and completely precluding deep architectural customization across diverse NICs. Second, despite bypassing the CPU for raw packet delivery, the end-to-end AI pipeline remains bound to host orchestration. Once data arrives in GPU memory, the CPU must still wake up to manage stateless buffers, track stateful telemetry, and explicitly launch the subsequent CUDA kernels for inference. This host-device coupling reintroduces the exact scheduling and batch-wait latency that direct I/O was intended to eliminate.  Third, these frameworks fail to bridge the semantic and microarchitectural mismatch between network traffic and AI execution, leaving data in VRAM as raw, asynchronous byte streams. Updating shared flow state and converting packets into AI-ready multi-dimensional tensors directly on a GPU triggers severe SIMT warp divergence and atomic contention, forcing developers to manually engineer custom synchronization and inference pipelines from scratch.

\section{The AGP Design: A Unified Vision}
To address the persistent CPU bottleneck of AI-integrated packet processing, AGP packages a closed-loop GPU execution environment into a reusable framework that owns the entire fast path.

\begin{figure}
    \centering
    \includegraphics[width=0.93\linewidth]{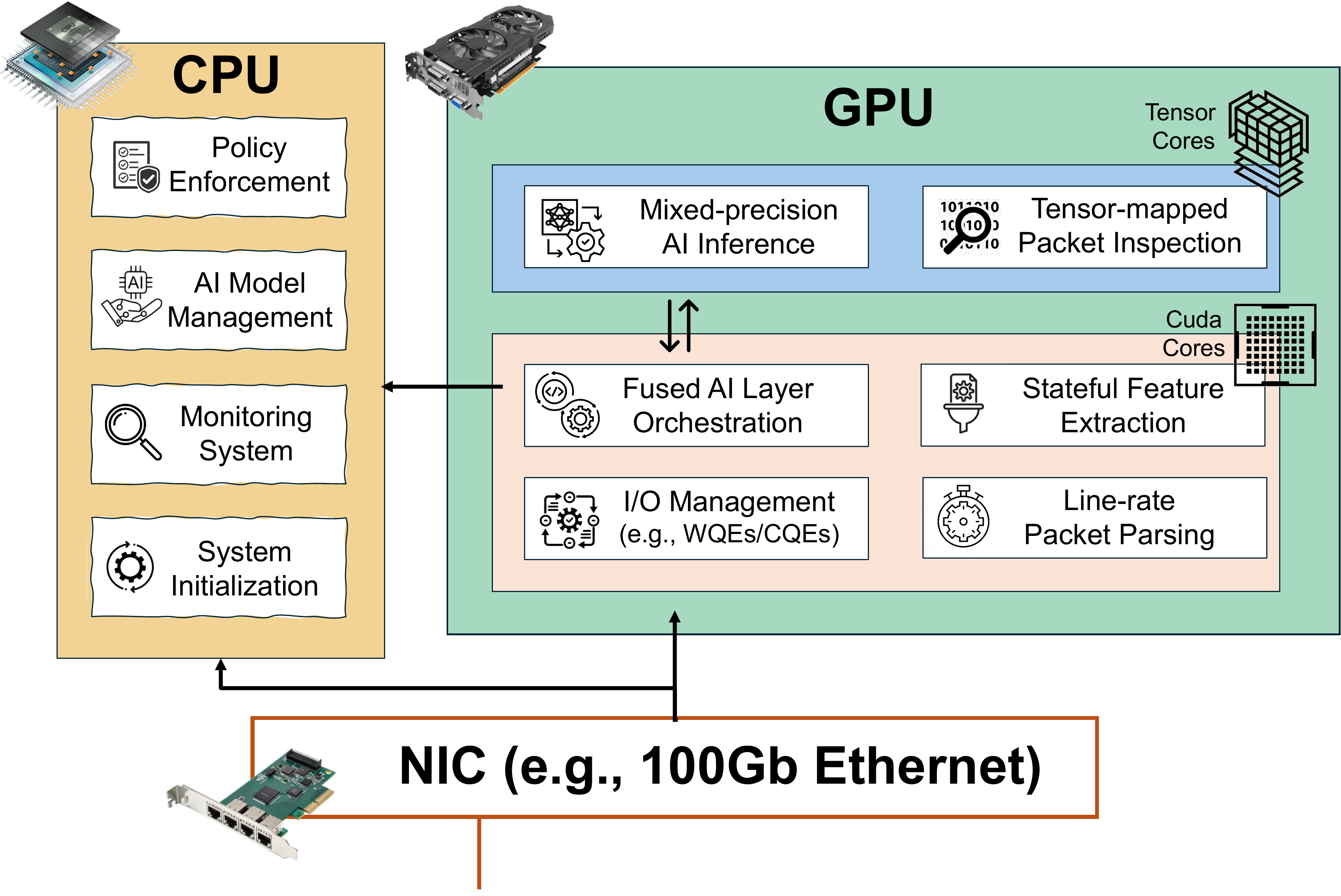}\vspace{-1mm}
    \caption{The Overview of AGP System Architecture.}
    \label{fig:agp_overview}
    \vspace{-3mm}
\end{figure}

\subsection{Overview}
Figure~\ref{fig:agp_overview} illustrates the end-to-end execution flow of AGP's bifurcated control-data plane architecture. AGP isolates all line-rate data processing within the GPU while restricting the CPU to asynchronous supervision off the critical path. During initialization, the CPU negotiates direct NIC-to-GPU memory access, allocates DMA rings (WQE/CQE) in GPU VRAM, and launches the persistent CUDA kernel driving the data plane. Thereafter, the CPU merely monitors system health, can hot-swap model weights into VRAM without stalling the pipeline, and enforces slow-path policy updates (e.g., routing or logging) upon receiving asynchronous control signals from the GPU.

The GPU, on the other hand, functions as an autonomous, closed-loop execution environment. Within the CUDA cores, a persistent I/O management stage continuously polls the DMA rings, directly ingesting packets and refreshing completion entries for the NIC. This raw byte stream is immediately passed to line-rate parsing, where parsing stages extract network telemetry and update hardware-accelerated stateful structures to maintain a live representation of traffic behavior. As flow features mature, dedicated CUDA warps normalize and reshape the data into multidimensional tensors. Rather than relying on a host-side AI framework, execution is managed by a persistent mega-kernel. This kernel autonomously orchestrates the deep learning pipeline, synchronizing data handoffs and managing the underlying hardware resources. Specifically, it can dispatch the heavy, mixed-precision matrix multiplications (GEMM)~\cite{micikevicius2018mixed, markidis2018nvidia} to the Tensor Cores for AI inference, while also leveraging them to execute tensor-mapped payload inspection (DPI) for complex pattern matching. By keeping the entire workflow from raw packet reception to final inference bound within device memory, AGP enables line-rate, stateful inline AI inference with zero host scheduling.

\subsection{Handing NIC Control to GPU}
True autonomy requires decoupling the control path from the CPU. DPDK's Poll Mode Driver dedicates a full core to each polled queue, so line-rate deployments often consume dozens of cores for I/O alone, making telemetry aggregation, preprocessing, and AI orchestration on this host-centric control plane prohibitively resource-intensive.

\BfPara{GPU-Native I/O Management (WQEs/CQEs)}  While the vendor-specific GPU networking frameworks (e.g., DOCA GPUNetIO) partially move network control onto the GPU, they remain proprietary and are maintained for specific hardware generations (e.g., ConnectX-7~\cite{CONNECTX7} and beyond), limiting extensibility and making it difficult to build systems. Upstream DPDK, to the best of our knowledge, stops at GPU-resident
payloads~\cite{dpdk_gpudev}. AGP removes both limits by placing the rings
themselves in VRAM using standard DPDK abstractions. Instead of placing NIC receive and transmit queues in host memory, AGP allocates these queues directly in GPU VRAM. During initialization, the host allocates queue memory in VRAM and establishes DMA mappings via I/O Virtual Addresses (IOVAs), allowing the NIC to issue transactions directly to GPU queues. Once mapped, the CPU is completely removed from the critical packet-processing path. Incoming descriptors and packet payloads are subsequently transferred directly into VRAM through PCIe peer-to-peer (P2P) DMA using the registered IOVA mappings. The persistent CUDA kernel then monitors queue state locally within GPU memory, enabling packet reception, queue management, and scheduling decisions to proceed without CPU intervention.

\BfPara{Enforcing DMA Memory Visibility} Unlike host memory, GPU memory lacks the cache-coherence assumptions of DPDK designs, which can lead to race conditions in which kernels observe queue updates before packet data becomes visible. AGP resolves this with cache-aware polling and device-wide synchronization: a polling thread accepts a CQE only after re-reading its ownership field following \texttt{\_\_threadfence\_system()}, and advances the consumer index only once the payload is observed, ensuring DMA updates are globally visible before processing begins.

\BfPara{Two Polling Modes for Line-rate Ingestion} With memory consistency enforced across device queues and DMA visibility, AGP exposes two execution modes for different semantic workloads. In the flow-level mode, AGP leverages the NIC's Receive Side Scaling (RSS) to distribute incoming traffic across multiple receive queues in VRAM. Then, each GPU warp is assigned to a specific queue, enabling massively parallel polling without contention. This execution model is particularly well-suited for workloads such as flow-level telemetry, traffic classification, and stateful feature extraction, allowing throughput to scale with the number of available GPU warps and streaming multiprocessors.

In the sequence-level mode, AGP handles workloads that require strict packet ordering or operate on a single logical stream. All packets arrive in a shared VRAM-resident ring buffer in NIC arrival order, and multiple warps cooperatively claim descriptors via lightweight atomic operations that preserve this order. Downstream kernels thus operate on temporally consistent streams, allowing sequence-dependent tasks such as inter-arrival time analysis, protocol state tracking, and AI inference over packet sequences to run without additional reordering or buffering.

\subsection{GPU-Native Stateful Telemetry} 

Once packets are ingested into VRAM, AGP bridges the semantic gap between
unstructured network byte streams and the rigidly formatted multidimensional
tensors required for AI inference. Replicating the CPU's branch-intensive
parsing and flow tracking on a parallel accelerator introduces severe
microarchitectural friction: protocol parsing induces warp divergence, while
concurrent state updates cause atomic contention. AGP therefore
bifurcates telemetry construction by workload semantics, ensuring feature
extraction uses strictly warp-aligned, divergence-minimizing memory operations.

\BfPara{Stateless Packet-Level Parsing}  For workloads without historical context (e.g., payload signature matching~\cite{wang2019hyperscan} or per-packet classification~\cite{holland2021new}), each packet is processed independently. In this execution path, threads within a warp perform inline feature extraction and feed the AI mega-kernel directly. Because there is no shared state across packets, this pipeline operates entirely lock-free, avoiding synchronization overhead and maximizing raw packet-processing throughput.

\BfPara{Stateful Flow-Level Aggregation} In contrast, flow-level telemetry requires shared state (e.g., connection counters, inter-arrival times, or sketch-based statistics). Because network traffic arrives in bursts, multiple threads within a warp frequently attempt to update overlapping flow identifiers simultaneously. Naively issuing global atomic operations (e.g., \texttt{atomicAdd}) to update shared hash tables leads to severe memory serialization. AGP eliminates this contention through a two-tiered, GPU-native memory hierarchy. First, inter-ring contention is eliminated via Receive Side Scaling (RSS), which deterministically hashes all packets of a given flow to the same VRAM receive ring. Second, intra-ring contention is resolved by replacing global atomic updates with warp-cooperative reductions. Using hardware shuffle primitives (e.g., \texttt{\_\_shfl\_down\_sync()}), threads aggregate per-flow statistics locally in high-speed registers before committing a single, coalesced update to memory.

\BfPara{Optimized Device Memory Layout} To minimize atomic pressure, AGP replaces pointer-heavy CPU data structures (e.g., linked lists) with flat, contiguous GPU arrays for all ring-local state. This ensures writes to global memory are coalesced, maximizing VRAM bandwidth utilization.

\begin{figure}[t]
    \centering
    \includegraphics[width=\linewidth]{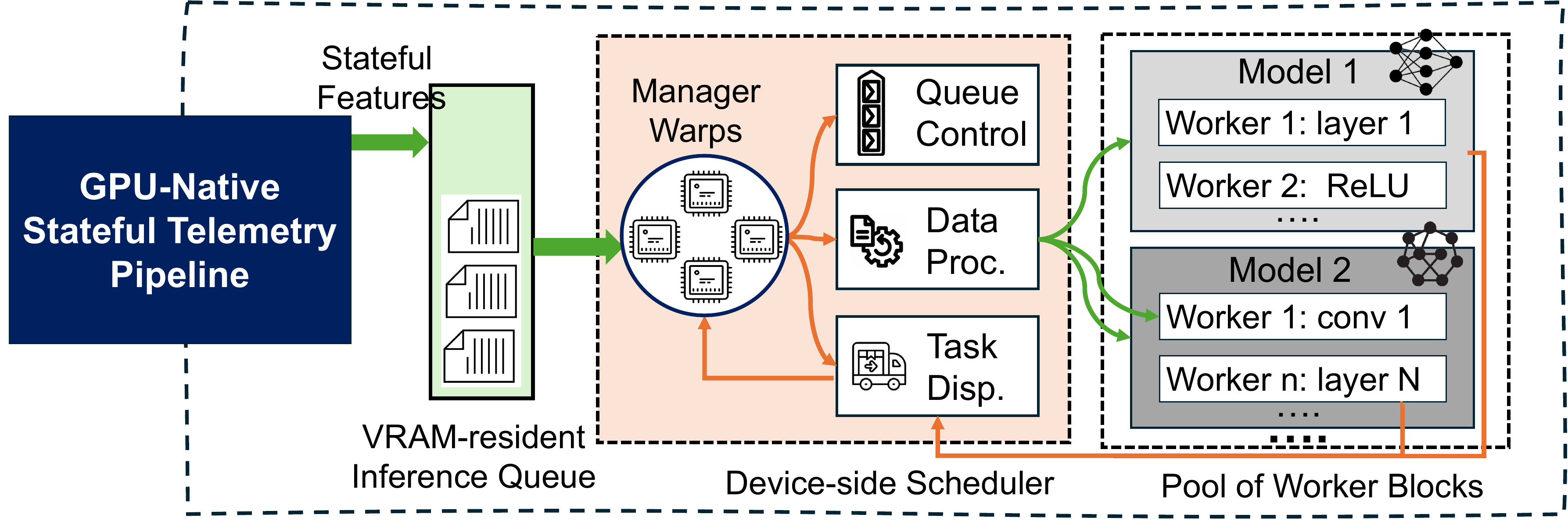}\vspace{-1mm}
    \caption{AGP’s GPU-Native Autonomous AI Inference.}
    \label{fig:AI_p}
    \vspace{-3mm}
\end{figure}

\subsection{The Autonomous AI Model Inference}
True data-path autonomy requires executing stateful downstream compute, such as multi-layer neural inference, directly on the GPU without host scheduling. In conventional frameworks (e.g., TensorFlow, PyTorch), decomposing models into dozens of CPU-launched micro-kernels introduces severe per-kernel synchronization overhead. Recent work~\cite{cheng2025mirage} mitigates this with persistent mega-kernels for Large Language Model (LLM) inference, but the approach does not transfer directly to the network environment.

First, in LLM serving, the CPU still feeds requests and collects outputs, whereas in-network processing already lives entirely within the GPU; the challenge is thus integrating inference into a high-throughput packet pipeline without
additional memory movement or synchronization. Second, existing mega-kernel designs are specialized for transformers and do not extend naturally to the lightweight convolutional models common in networking tasks. AGP therefore requires an execution model that tightly couples packet processing with inference while remaining flexible across model architectures.

\BfPara{Fused Single-Block Execution} For lightweight models, AGP bypasses inter-layer scheduling via aggressive kernel fusion. The entire inference pipeline executes within a single CUDA thread block. By keeping all intermediate activations strictly on-chip (e.g., shared memory/L1) and using local synchronization (\texttt{\_\_syncthreads()}), this eliminates global memory traffic and launch overhead. 

\BfPara{Device-Side Scheduling and Execution Queues} For multi-stage models that exceed the register and shared memory limits of a single streaming multiprocessor (SM), AGP introduces a hierarchical, device-side scheduling engine. Instead of a monolithic block, execution is dynamically coordinated across multiple worker blocks. Once the telemetry pipeline finalizes the stateful features, the data is pushed into a VRAM-resident inference queue. A dedicated set of manager warps continuously monitors this queue. When a telemetry tensor is ready, these managers handle final preprocessing (e.g., precision casting or dimensionality reduction) and dynamically dispatch the data to idle worker blocks assigned to compute specific model layers. As worker blocks complete their assigned task (e.g., convolution, activation), they publish their execution state through lock-free, GPU-resident synchronization flags. The manager threads track this state graph in real time, resolving layer dependencies and triggering subsequent layers. This closed-loop scheduling ensures that even complex, multi-stage inferences progress layer-by-layer entirely within the device, maintaining line-rate throughput without host intervention, as shown in Figure~\ref{fig:AI_p}.

\begin{figure*}[t]
    \centering 
    \subfigure[AGP (Ours)]{\includegraphics[width=0.32\textwidth]{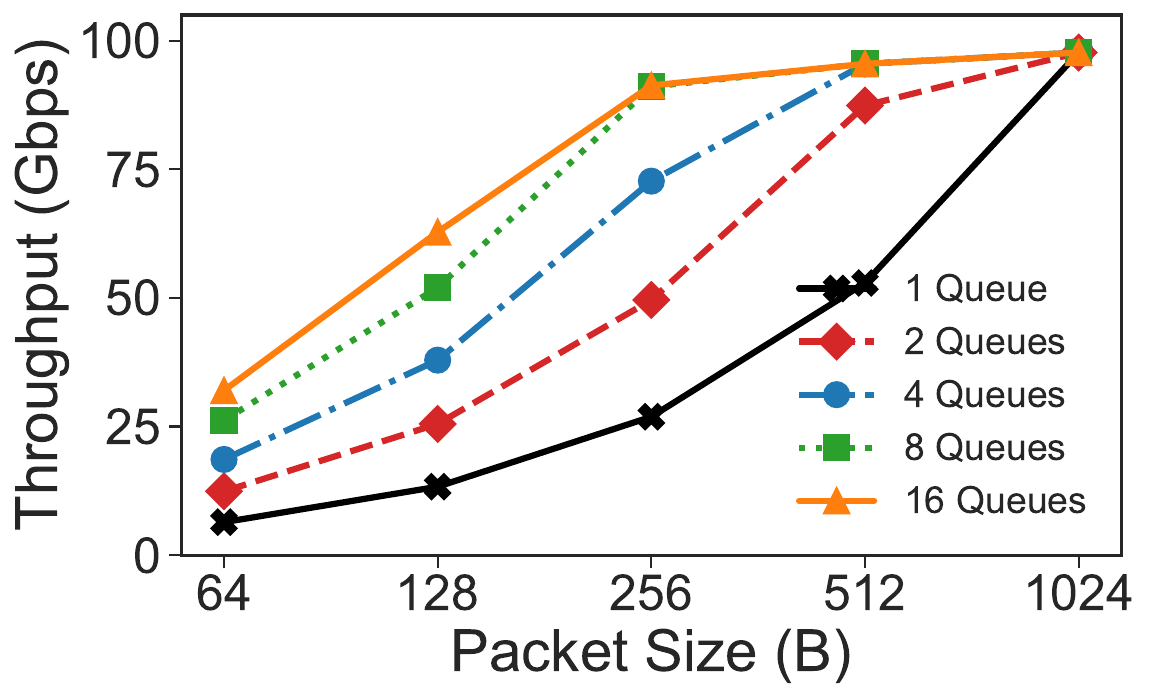}}\hfill
    \subfigure[CPU / DPDK]{\includegraphics[width=0.32\textwidth]{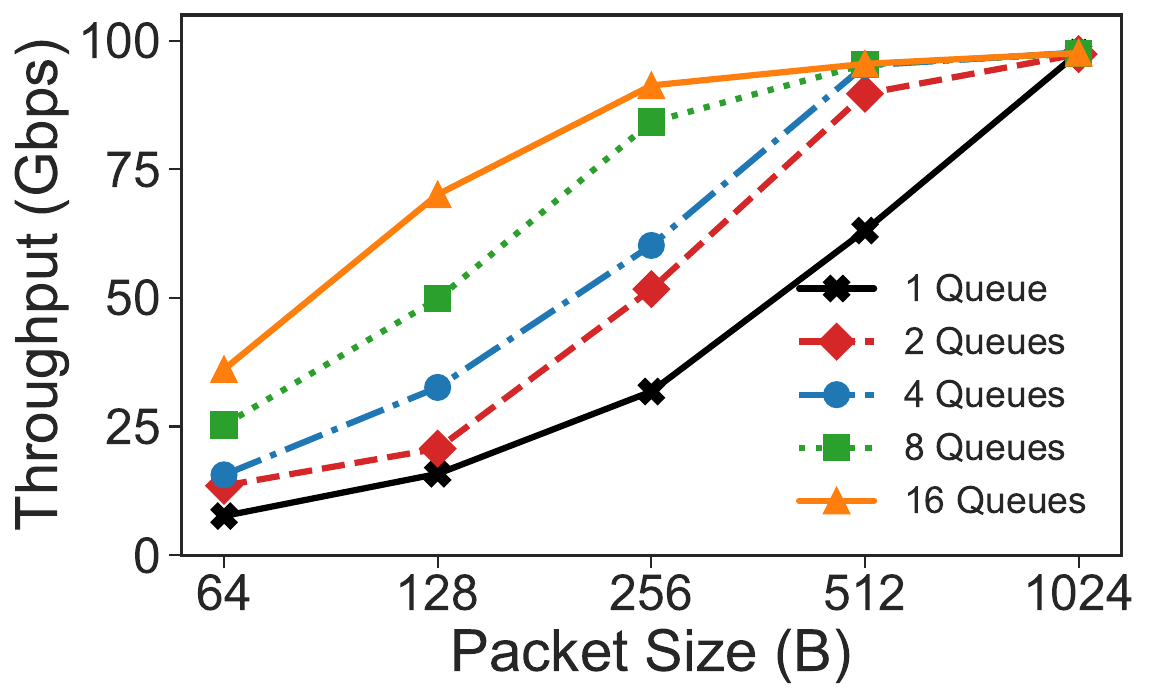}}\hfill
    \subfigure[DOCA]{\includegraphics[width=0.32\textwidth]{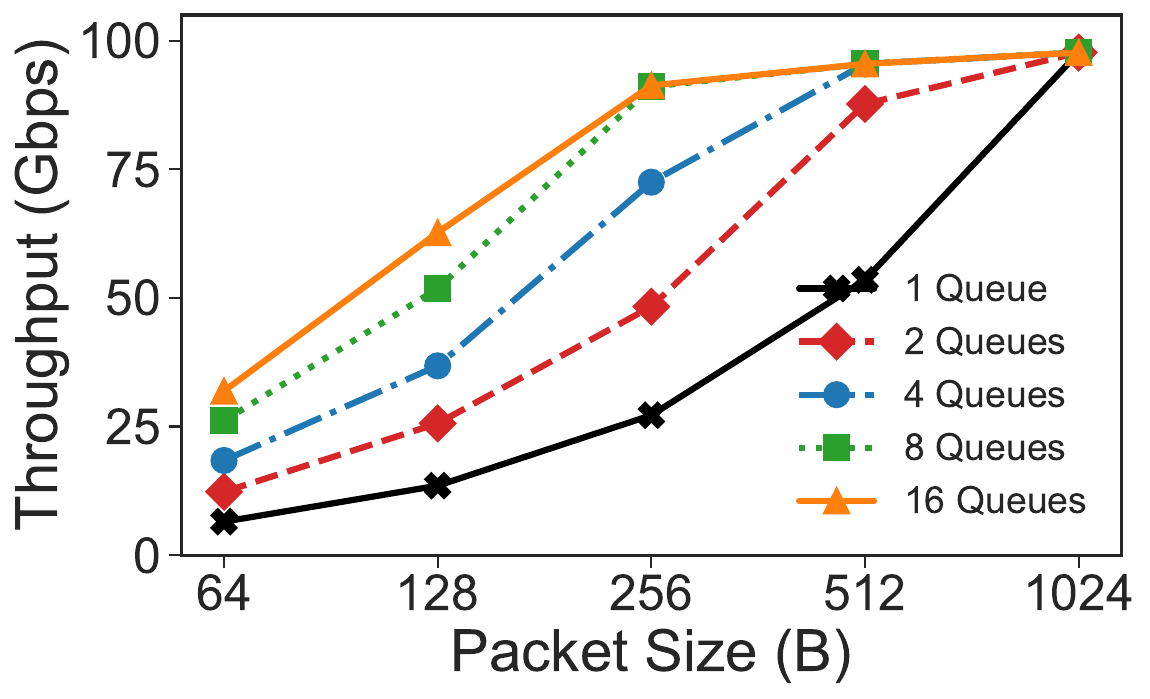}}
    \vspace{-1mm}
    \caption{System throughput scaling across varying packet sizes and RX queues for (a) AGP, (b) CPU/DPDK, and (c) DOCA baselines. All systems successfully saturate the 100 Gbps network limit for packets 512 bytes and larger.}
    \label{fig:CMP}\vspace{-3mm}
\end{figure*}

\section{Case Study}
We evaluate AGP using SketchFeature~\cite{kim2025sketchfeature}, a state-of-the-art intrusion detection system (IDS), as an end-to-end stress test for autonomous GPU data-path control. Unlike legacy GPU packet-processing systems~\cite{Hanetal2010, Vasiliadisetal2011, Jamshedetal2012} that rely on stateless per-packet inspection or coarse CPU batching, SketchFeature exercises all stages of AGP, including packet processing, shared state management, dynamic feature extraction, and data-dependent GPU inference across thousands of concurrent flows. While traditional and CPU-bypass designs must fall back to the host CPU to coordinate state and schedule inference, AGP executes and schedules this entire closed-loop state machine autonomously inside the GPU.

\BfPara{Baselines} We benchmark AGP against two baselines representing current network-to-AI integration: 1) \textit{CPU-Mediated DPDK:} The NIC DMAs packets into host DRAM, where dedicated DPDK Poll Mode Driver (PMD) cores parse traffic and extract stateful telemetry; two additional CPU cores construct multidimensional tensors, push them across PCIe, and launch PyTorch CUDA kernels.
2) \textit{NVIDIA DOCA:} Since native DOCA restricts support to newer NIC hardware, we
run NVIDIA's GPUNetIO GPU-side receive code unmodified and replace only the
queue setup with our DPDK library to establish direct peer-to-peer (P2P) DMA from the NIC to VRAM, with the sketch also kept in VRAM. We dedicate one CPU core in the control plane to build the tensors and launch PyTorch kernels (we observed adding cores raises inference capacity by 1.5$\times$ but increases latency by 76\%), isolating the tax of host-side scheduling when raw data movement is already optimized.
All three systems run the same CNN classifier from
SketchFeature~\cite{kim2025sketchfeature}. AGP classifies each flow using a single warp within a persistent mega-kernel, whereas both baselines use the LibTorch C++ API (PyTorch 2.10)~\cite{paszke2019pytorch} with kernel fusion and CUDA Graphs explicitly enabled, the two optimizations that PyTorch's \texttt{max-autotune} mode applies, which is standard practice
for minimizing scheduling overhead.

\BfPara{Hardware Testbed} Our prototype runs on a dual-socket Intel Xeon Platinum 8358 server (2$\times$32 cores, 2.6 GHz) with 1 TB of host DRAM, an NVIDIA A100 GPU, and two Mellanox ConnectX-6 100 GbE NICs connected via PCIe Gen4 $\times$16. A custom DPDK traffic generator on one NIC transmits packets to the second NIC colocated with the A100 GPU.

\subsection{System Throughput and Efficiency}
Figure~\ref{fig:CMP} presents a comprehensive evaluation of system throughput, detailing performance across both varying packet sizes and RX queue configurations for all three architectures. Under extreme stress with synthetic 64-byte packets, the CPU/DPDK baseline achieves a slightly higher maximum packet rate (75.2 Mpps vs. 66.6 Mpps). This gap is due to the NIC-to-GPU PCIe path, where it was observed that the NIC discards packets it cannot deliver to GPU memory across the CPU root complex. We note that none of the systems reaches the theoretical 148.8 Mpps; the CPU/DPDK ceiling of 75.2 Mpps is set by our traffic generator's maximum transmit rate. From 512-byte packets onward, all three architectures saturate the 100 Gbps link (95.5 Gbps) with four or more RX queues. Most importantly, this demonstrates that AGP's autonomous in-GPU pipeline matches the raw networking performance of heavily multi-cored, host-centric designs across all realistic traffic profiles, scaling its throughput seamlessly as more RX queues are utilized. 
Beyond static line-rate processing, we also verified system resilience under dynamic traffic patterns; when tested for a real-world CAIDA~\cite{CAIDA} trace workload, AGP successfully sustained up to 700k new flows per second (83.5 Gbps) with zero packet loss and classification of all detectable flows. This confirms that in-GPU state management safely absorbs extreme flow concurrency without bottlenecking the fast path.
Table~\ref{tab:power} reports power consumption using 16 RX queues. We must acknowledge that deploying a datacenter-class GPU introduces a substantial baseline idle power draw ($\approx$61.5 W). However, when evaluating the \textit{marginal} power cost of aggressive network polling and line-rate feature extraction, we observe a clear architectural advantage. While the CPU/DPDK pipeline requires $121$ W of extra CPU power to saturate throughput, AGP requires merely $0.4$ W of additional host CPU power and utilizes just $20.1$ W of active GPU computational overhead. Furthermore, keeping even a single CPU core active at its maximum frequency for CPU/DPDK or DOCA processing prevents the socket from entering package sleep, incurring a hidden $\approx$28-60 W uncore and memory fabric penalty. Because AGP completely bypasses host polling, it avoids this shared socket cost, highlighting its superior marginal power efficiency.

\begin{table}[t]
\centering
\caption{Average Marginal Power Overhead.}\vspace{-3mm}
\label{tab:power}
\resizebox{0.49\textwidth}{!}{
\begin{tabular}{lccc}
\toprule
\textbf{Metric} & \textbf{AGP (Ours)} & \textbf{CPU/DPDK} & \textbf{DOCA} \\
\midrule
Marginal CPU (W) & $0.4$ & $121$ & $7.6$ \\
Marginal GPU (W) & $20.1$ & $5.1$ & $20.8$ \\
\midrule
\textbf{Total Overhead (W)} & $\mathbf{20.5}$ & $\mathbf{126.1}^{*}$ & $\mathbf{28.4}^{*}$ \\
\bottomrule
\multicolumn{4}{l}{\footnotesize \textit{*Excludes an additional $\approx$28-60 W socket wake penalty (uncore/fabric).}}
\end{tabular}
}\vspace{-3mm}
\end{table}

\subsection{End-to-End Latency and Inference}
\begin{figure}
    \centering
    \includegraphics[width=0.43\textwidth]{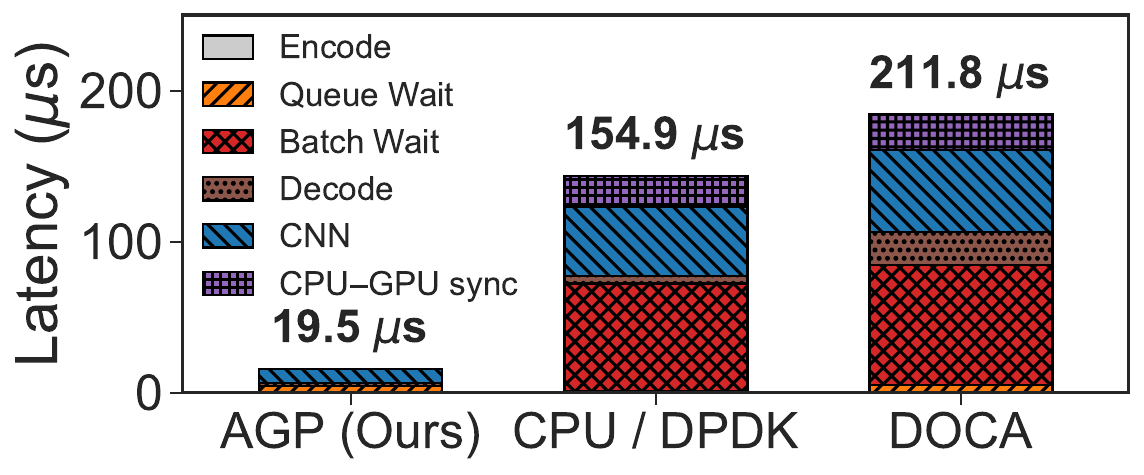}\vspace{-1mm}
    \caption{End-to-end latency breakdown. Both baselines are fully optimized using kernel fusion and CUDA Graphs.}
    \label{ig:latency_breakdown}
    \vspace{-2mm}
\end{figure}

To understand the execution bottlenecks of each architecture, we break down the end-to-end performance of the flow classification pipeline, capturing the time from the arrival of the packet that triggers inference to the final inference score. As shown in Figure~\ref{ig:latency_breakdown} under saturation (line-rate traffic with a low inference threshold, so inference is triggered
continuously), AGP achieves a median end-to-end latency of $19.5\,\mu\text{s}$, compared with the DPDK ($154.9\,\mu\text{s}$) and DOCA ($211.8\,\mu\text{s}$) baselines. This difference comes directly from the host-side scheduling overheads. The DOCA pipeline explicitly highlights the illusion of autonomy: although packets land in VRAM, CPU-mediated kernel launches introduce severe synchronization penalties, including extensive queue/batch wait times and a lengthy CNN execution time. Conversely, AGP's persistent mega-kernel removes all host-launch gaps and batching, reducing queue wait to $4.3\,\mu\text{s}$ and CNN execution to just $9.28\,\mu\text{s}$ per flow.

These synchronization bottlenecks severely cap inference throughput under heavy load. Table~\ref{tab:Inference} compares isolated AI capacity with the saturation limits of the whole system. While the CPU-launched DOCA baseline achieves 2.1M classifications per second in an AI-only vacuum (Inference Path), it degrades to 0.57M flows per second when running concurrently with packet processing. The CPU/DPDK pipeline similarly collapses to 0.15M. Conversely, AGP's persistent mega-kernel completely bypasses host limits, reaching 6M in isolation and maintaining 5.3M flows per second under full system saturation. This $9.3\times$ gain factors into $2.8\times$ from the kernel itself (Model Only) and a further $3.3\times$ from removing host scheduling. AGP achieves its peak throughput using just $\sim$2\% of the GPU's total thread capacity (144 warps of 32 threads vs. the A100's $\approx$221k), whereas each PyTorch batch in the baselines consumes over 72\% of the total thread capacity, yet the CPU-driven kernel orchestration significantly degrades flow inference throughput.

Beyond speed, AGP frees the host entirely: it occupies no host CPU cores, whereas DOCA requires a dispatcher core, and CPU/DPDK requires 18 (16 RX cores, a dispatcher, and a sketch-maintenance core). At realistic loads, all three systems keep up due to the mix of packet sizes, so AGP's benefit is not throughput but latency and freed host cores. At 60k new flows per second, a typical cloud arrival rate~\cite{liu2024serveflow}, replaying CAIDA flows, all three systems classify every detectable flow, yet AGP delivers a median/p99 latency of $19/24\,\mu\text{s}$ versus $438/776\,\mu\text{s}$ for CPU/DPDK and $479/857\,\mu\text{s}$ for DOCA, where slower arrivals make batch-wait time dominate. 

\begin{table}[t]
\centering
\caption{End-to-End Inference Throughput (Flows/s).}\vspace{-3mm}
\label{tab:Inference}
\resizebox{0.49\textwidth}{!}{
\begin{tabular}{l c c c}
\toprule
\textbf{Metric} & \textbf{AGP (Ours)} & \textbf{CPU/DPDK} & \textbf{DOCA} \\
\midrule
Model Only (CNN) & $\mathbf{11.6M}$ & $4.1M$ & $4.1M$ \\
Inference Path (Decode + CNN) & $\mathbf{6.0M}$ & $0.53M$ & $2.1M$ \\
Whole System (with NIC) & $\mathbf{5.3M}$ & $0.15M$ & $0.57M$ \\
\bottomrule
\end{tabular}
}\vspace{-3mm}
\end{table}

\subsection{System Footprint}
Table~\ref{tab:footprint} summarizes the resources each architecture occupies with 16 RX queues. AGP runs without any host CPU involvement: its single persistent kernel
uses 32 thread blocks (16 RX warps and 128 inference warps), i.e., at most 32 of the A100's 108 SMs, leaving the remaining GPU capacity available for co-located workloads.
In contrast, CPU/DPDK dedicates 18 host cores to polling, dispatching, and sketch maintenance, and both baselines carry the host memory footprint of the PyTorch runtime.
DOCA occupies the largest share of VRAM, as it keeps both the packet buffers and the sketch in GPU memory in addition to the PyTorch runtime, yet still depends on a
dedicated host core to drive inference.

\begin{table}[t]
\centering
\caption{System footprint with 16 RX queues.}\vspace{-3mm}
\label{tab:footprint}
\resizebox{0.49\textwidth}{!}{
\begin{tabular}{lccc}
\toprule
\textbf{Resource} & \textbf{AGP (Ours)} & \textbf{CPU/DPDK} & \textbf{DOCA} \\
\midrule
Dedicated CPU cores & $0$ & $18$ & $1$ \\
Host DRAM (total) & $0.35$ GB & $6.8$ GB & $5.4$ GB \\
GPU VRAM$^{\ddagger}$ & $2.4$ GB & $1.4$ GB & $3.4$ GB \\
Persistent GPU warps (RX / Inference) & $16$ / $128$ & $0$ / 0 & $16$ / 0 \\
GPU kernel launches & $1$ (once) & per batch & per batch \\
GPU warps per inference batch (32 / 256 flows) & - & 5k / 16k & 5k / 16k \\
\bottomrule
\end{tabular}
}
{\footnotesize $^{\ddagger}$Includes the CUDA context.} \vspace{-3mm}
\end{table}

\section{Discussion \& Opportunities}

\BfPara{Practical Implications} Our prototype's residual throughput gap at 64-byte packets stems from a sub-optimal PCIe topology that forces P2P traffic through the CPU root complex. Platforms where the NIC and GPU share a PCIe switch, or are integrated on one device, should close this gap. In our prototype, we targeted the mlx5 driver; porting to other NIC families is left as future work.

\BfPara{GPU as a Multi-Tenant Network Endpoint} Beyond IDS, because AGP's entire pipeline
fits in a small, persistent slice of SMs, that slice can serve as a network
front-end for the rest of the GPU. Remote users stream compute requests over
the NIC, and on-device manager warps authenticate, parse, and dispatch them to
tenant kernels on the remaining SMs without host OS intervention. This requires
GPU-native control: with a host in the loop, every request pays the same
launch and batch-wait latency we measured for the baselines. Similarly, it can be extended to inline semantic request routing for distributed AI serving, and telemetry and congestion control co-located with primary training jobs.

\BfPara{A Unified AI-Network Compiler} While AGP demonstrates the benefits of autonomous GPU networking, developing fused GPU kernels and hardware-aware memory layouts remains complex and error-prone. A promising future direction is a unified compiler that bridges networking and AI, replacing today’s fragmented toolchain (e.g., DPDK, CUDA, and PyTorch). Such a compiler could jointly compile packet processing and AI model graphs into optimized GPU kernels and memory layouts, enabling programmable, stateful AI data planes with minimal developer effort.

\section{Conclusion}
In this paper, we introduced AGP, a GPU-native framework for autonomous packet processing that decouples the network data path from the host CPU. By leveraging an in-GPU execution environment and a fused mega-kernel pipeline, AGP provides a unified AI-native execution environment with zero host intervention, achieving line-rate packet processing with $6.1\times$ lower marginal power consumption than CPU/DPDK and accelerating network AI inference by $9.3\times$ over host-scheduled bypass baselines, providing a foundation for next-generation intelligent network infrastructure. More broadly, once the GPU owns the data path, per-flow scheduling, in-place state, and a host-free fast path follow, making the GPU a first-class network endpoint.

\begin{acks}
This material is based upon work supported by the National Science Foundation under Grant No.~2542128.
\end{acks}

\bibliographystyle{ACM-Reference-Format} 
\bibliography{references}

\end{document}